\documentclass[3p,times]{elsarticle}

\usepackage{ecrc}

\volume{00}

\firstpage{1}

\journalname{Procedia Computer Science}

\runauth{}

\jid{procs}

\jnltitlelogo{Procedia Computer Science}

\CopyrightLine{2011}{Published by Elsevier Ltd.}

\usepackage{amssymb}

\usepackage[figuresright]{rotating}

\begin{document}

\begin{frontmatter}



\dochead{}

\title{Recent highlights from ARGO-YBJ}


\author{Di Sciascio Giuseppe on behalf of the ARGO-YBJ Collaboration}

\address{INFN - Sezione Roma Tor Vergata, Roma, Italy}

\begin{abstract}
The ARGO-YBJ experiment has been in stable data taking for 5 years at the YangBaJing Cosmic Ray Observatory (Tibet, P.R. China, 4300 m a.s.l., 606 g/cm$^2$). With a duty-cycle greater than 86\% the detector collected about 5$\times $10$^{11}$ events in a wide energy range, from few hundreds GeV up to the PeV.
A number of open problems in cosmic ray physics has been faced  exploiting different analyses.
In this paper we summarize the latest results in gamma-ray astronomy and in cosmic ray physics.
\end{abstract}


\end{frontmatter}


\section {The ARGO-YBJ experiment}
\label{}

ARGO-YBJ is a full coverage air shower detector located at the Yangbajing Cosmic Ray Observatory (Tibet, PR China, 4300 m a.s.l., 606 g/cm$^2$) devoted to the study of gamma rays and cosmic rays.
Exploiting the high altitude and the full coverage technique, ARGO-YBJ can detect gamma rays with an energy threshold  as low as a few hundreds GeV.

The detector is constituted by a central carpet $\sim$74$\times$78 m$^2$, made of a single layer of resistive plate chambers (RPCs) with $\sim$93$\%$ of active area, enclosed by a guard ring partially instrumented ($\sim$20$\%$) up to $\sim$100$\times$110 m$^2$. The apparatus has a modular structure, the basic data acquisition element being a cluster (5.7$\times$7.6 m$^2$), made of 12 RPCs (2.85$\times$1.23 m$^2$ each). Each chamber is read by 80 external strips of 6.75$\times$61.80 cm$^2$ (the spatial pixels), logically organized in 10 independent pads of 55.6$\times$61.8 cm$^2$ which represent the time pixels of the detector \cite{aielli06}. 
The readout of 18,360 pads and 146,880 strips is the experimental output of the detector. 
In addition, in order to extend the dynamical range up to PeV energies, each chamber is equipped with two large size pads (139$\times$123 cm$^2$) to collect the total charge developed by the particles hitting the detector \cite{bigpad}.
The RPCs are operated in streamer mode by using a gas mixture (Ar 15\%, Isobutane 10\%, TetraFluoroEthane 75\%) for high altitude operation \cite{bacci00}. The high voltage settled at 7.2 kV ensures an overall efficiency of about 96\% \cite{aielli09a}.
The central carpet contains 130 clusters and the full detector is composed of 153 clusters for a total active surface of $\sim$6,700 m$^2$. The total instrumented area is $\sim$11,000 m$^2$.
For each event the location and timing of every detected particle is recorded, allowing the reconstruction of the lateral distribution and the arrival direction. A simple, yet powerful, electronic logic has been implemented to build an inclusive trigger. This logic is based on a time correlation between the pad signals depending on their relative distance. In this way, all the shower events giving a number of fired pads N$_{pad}\ge$ N$_{trig}$ in the central carpet in a time window of 420 ns generate the trigger. This trigger can work with high efficiency down to N$_{trig}$ = 20, keeping negligible the rate of random coincidences.

Because of the small pixel size, the detector is able to record events with a particle density exceeding 0.003 particles m$^{-2}$, keeping good linearity up to a core density of about 15 particles m$^{-2}$.
This high granularity allows a complete and detailed three-dimensional reconstruction of the front of air showers at an energy threshold of a few hundreds GeV. Showers induced by high energy primaries ($>$ 100 TeV) are also imaged by the charge readout of the large size pads \cite{bigpad}.

The whole system has been in stable data taking from November 2007 to January 2013, with the trigger condition N$_{trig}$ = 20 and a duty cycle $\geq$86\%. The trigger rate is $\sim$3.5 kHz with a dead time of 4$\%$.

Details on the analysis procedure (e.g., reconstruction algorithms, data selection, background evaluation, systematic errors) are discussed in \cite{aielli10,bartoli11a,bartoli11b}.
The performance of the detector (angular resolution, pointing accuracy, energy scale calibration) and the operation stability are continuously monitored by observing the Moon shadow, i.e., the deficit of CRs detected in its direction \cite{bartoli11a,bartoli12a}. 
ARGO-YBJ observes the Moon shadow with a sensitivity of $\sim$9 standard deviations (s.d.) per month. 
The measured angular resolution is better than 0.5$^{\circ}$ for CR-induced showers with energy E $>$ 5 TeV and the overall absolute pointing accuracy is $\sim$0.1$^{\circ}$.
The absolute pointing of the detector is stable at a level of 0.1$^{\circ}$ and the angular resolution is stable at a level of 10\% on a monthly basis.
The absolute rigidity scale uncertainty of ARGO-YBJ is estimated to be less than 13\% in the range 1 - 30 TeV/Z \cite{bartoli11a,bartoli12a}.

In this paper the latest results obtained in gamma-ray astronomy and in cosmic ray (CR) physics are summarized.

\section{Northern Sky Survey at TeV photon energies}

The ARGO-YBJ data used in this analysis were collected from November 2007 to January 2013, with a total observation time of 1670.45 days.  The total number of events selected with a zenith angle less than 50$^{\circ}$ is about 3$\times$10$^{11}$. They are used to fill a map in celestial coordinates (right ascension and declination) with $0.1^{\circ}\times0.1^{\circ}$ bins, covering the declination band from -10$^{\circ}$ to 70$^{\circ}$.

\begin{figure}
\centerline {\includegraphics[width=\textwidth,height=2.3in]{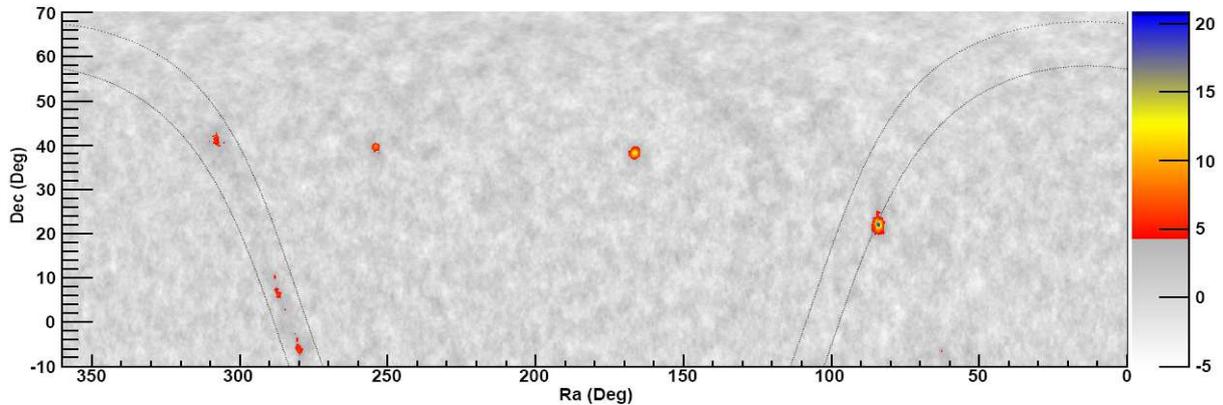}}
\caption{Significance map of the northern sky as seen by the ARGO-YBJ experiment in VHE $\gamma$-ray band.   The two dotted lines indicate the Galactic latitudes $b=\pm5^{\circ}$. The color scale shows the statistical significance in standard deviations (s.d.).}
\label{fig:fig01}
\end{figure}
%
\begin{figure}
\centerline {\includegraphics[width=\textwidth,height=2.3in]{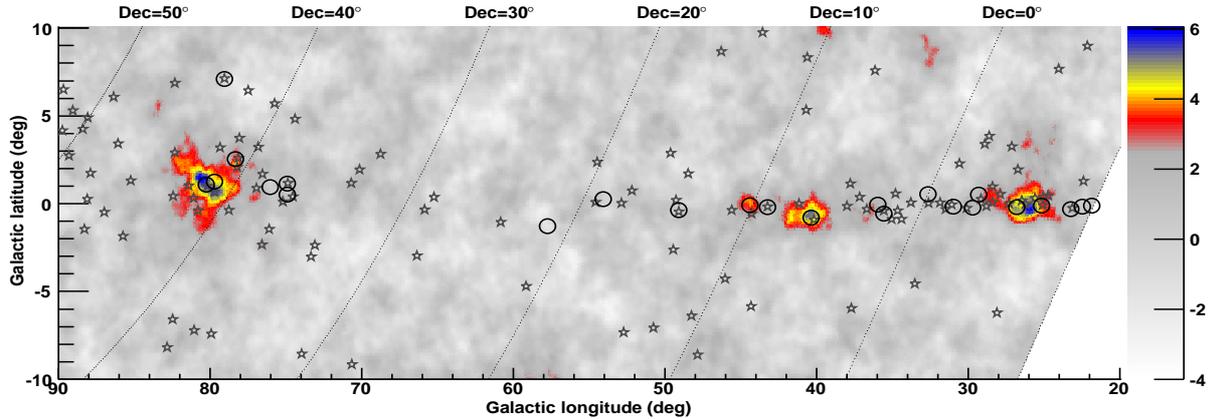}}
\caption{Significance map of the Galactic Plane region with $\mid b\mid<10^{\circ}$ and $20^{\circ}<l<90^{\circ}$ 
obtained by the ARGO-YBJ experiment.  The circles indicate the position of all the known VHE sources.  The open stars mark the location of the GeV sources of the second $Fermi$-LAT catalog.}
\label{fig:fig02}
\end{figure}
%

The significance distribution of the whole map bins, with a mean value of 0.002 and $\sigma$=1.02, closely follows a standard Gaussian distribution except for a tail with large positive values, due to the excesses from several gamma ray emission regions, shown in Fig. \ref{fig:fig01} \cite{bartoli13a}.
Table 1 lists the locations of all the regions with pre-trial significance greater than 4.5 s.d. . 
For each independent region, only the coordinates of the pixel with the highest significance are given.
Based on the distribution of negative points, a significance threshold of 4.5 s.d. corresponds to $\sim$ 2 false sources in our catalog.

Fig. \ref{fig:fig02} shows the inner Galactic Plane region longitude ($20^{\circ}<l<90^{\circ}$ and latitude $\mid b\mid<2^{\circ}$) in galactic coordinates. The locations of all the excesses with a significance greater than 4.0 s.d. are listed in Table 1.
Since the smoothing radius is larger than the bin width, the significances in adjacent bins are correlated, and a  Monte Carlo simulation is necessary to correctly evaluate the post-trial probabilities.
According to our simulations, the probability to have at least one source with a pre-trial significance $>$ 5.1 $\sigma$ ($>$ 4.0 $\sigma$) anywhere in the map (in the inner Galactic Plane) due to background fluctuations is 5$\%$.
However, since in the sky region monitored by ARGO-YBJ only $\sim$70 known VHE emitters exist, the post-trial significance increases for the candidate sources associated to a counterpart.
Details about different sources are discussed in \cite{bartoli13a}.

\begin{table}
\begin{center}
\caption{Location of the excess regions}
\begin{tabular}{cccccccc}
\hline
 ARGO-YBJ& Ra  &  Dec  &      S   &  Associated   \\
Name &   (deg)  &  (deg)  &       (s.d.)  &   TeV  Source \\
\hline
  J0409$-$0627&62.35  &   -6.45  &  4.8 & \\
  J0535+2203&83.75  & 22.05  & 20.8  & Crab Nebula \\
  J1105+3821&166.25  & 38.35    &14.1 & Mrk 421 \\
  J1654+3945&253.55  & 39.75    & 9.4 & Mrk 501 \\
  J1839$-$0627&279.95  & -6.45    & 6.0 & HESS J1841-055\\
  J1907+0627&286.95  & 6.45 & 5.3  &HESS J1908+063 \\
  J1910+0720&287.65  & 7.35  & 4.3  & \\
  J1912+1026&288.05  & 10.45  & 4.2  &HESS J1912+101\\
  J2021+4038&305.25  & 40.65  & 4.3  & VER J2019+407 \\
  J2031+4157&307.95    & 41.95   & 6.1    &MGRO J2031+41\\
 & & & &   TeV J2032+4130\\
 J1841-0332&280.25    & -3.55   & 4.2    & \\
\hline
\end{tabular}
\end{center}
 \vspace{-3mm}
\end{table}

\subsection{Sky upper limits}

Excluding the sources listed in Table 1, we can set upper limits to the $\gamma$-ray flux from all other directions in the sky. To estimate the  response of the ARGO-YBJ detector we simulated a source located at different declinations, with a power law spectrum in the energy range 10 GeV - 100 TeV and different spectral indices. The number of events is transformed  into a flux using the results of the simulation. The 95\% C.L. upper limits to the flux of $\gamma$-rays with energies above 500 GeV for each bin are obtained \cite{bartoli13a}.

\begin{figure}
\centerline{\resizebox{70mm}{!} {\includegraphics{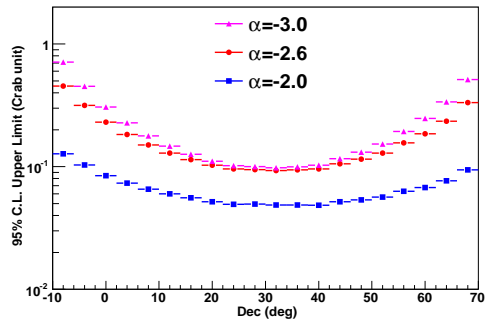}}}
\caption[h]{95\% C.L. flux upper limits for energy above 500 GeV, averaged over the right ascension, as a function of the declination. The different curves indicate a different power-law spectral index.}
\label{fig:fig03}
\end{figure}

The upper limits as a function of the declination are shown in Fig. \ref{fig:fig03} for different photon spectral indices. 
The limits range  between  9\% and  44\% $I_{Crab}$ and are the lowest obtained so far.
The lowest limit for a spectral index $-$2.0 ($-$3.0) is 5\% (9\%) $I_{Crab}$, where the Crab unit is defined as 5.77$\times$10$^{-11}$ cm$^{-2}$ s$^{-1}$ .
%
%

\subsection{The Fermi Cocoon in the Cygnus region}

\begin{figure}
\begin{minipage}[ht]{.48\linewidth}
  \centerline{\includegraphics[width=\textwidth]{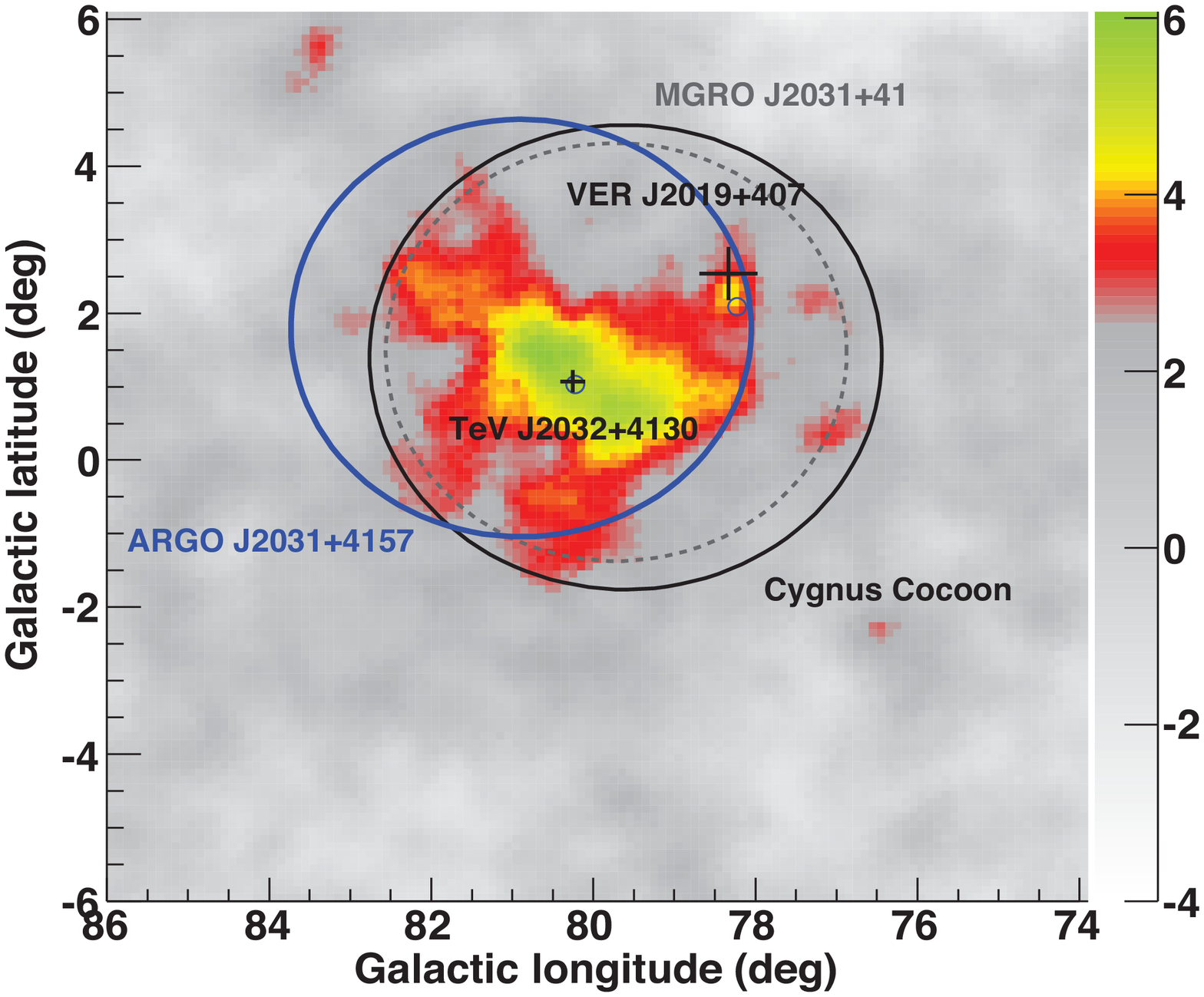}}
    \caption{The significance map of the ARGO J2031+4157 region observed by the ARGO-YBJ experiment. The 68\% containment circles for ARGO J2031+4157, MGRO J2031+41 and Cygnus Cocoon are shown. The position and extension of TeV 2032+4130 and VER J2019+407 are marked with crosses \cite{aharonian05, aliu11, aliu13}. The small circles indicate the positions of PSR 2021+4026 and PSR 2032+4127.}
\label{fig:fig04}
\end{minipage}\hfill
\begin{minipage}[ht]{.48\linewidth}
  \centerline{\includegraphics[width=\textwidth]{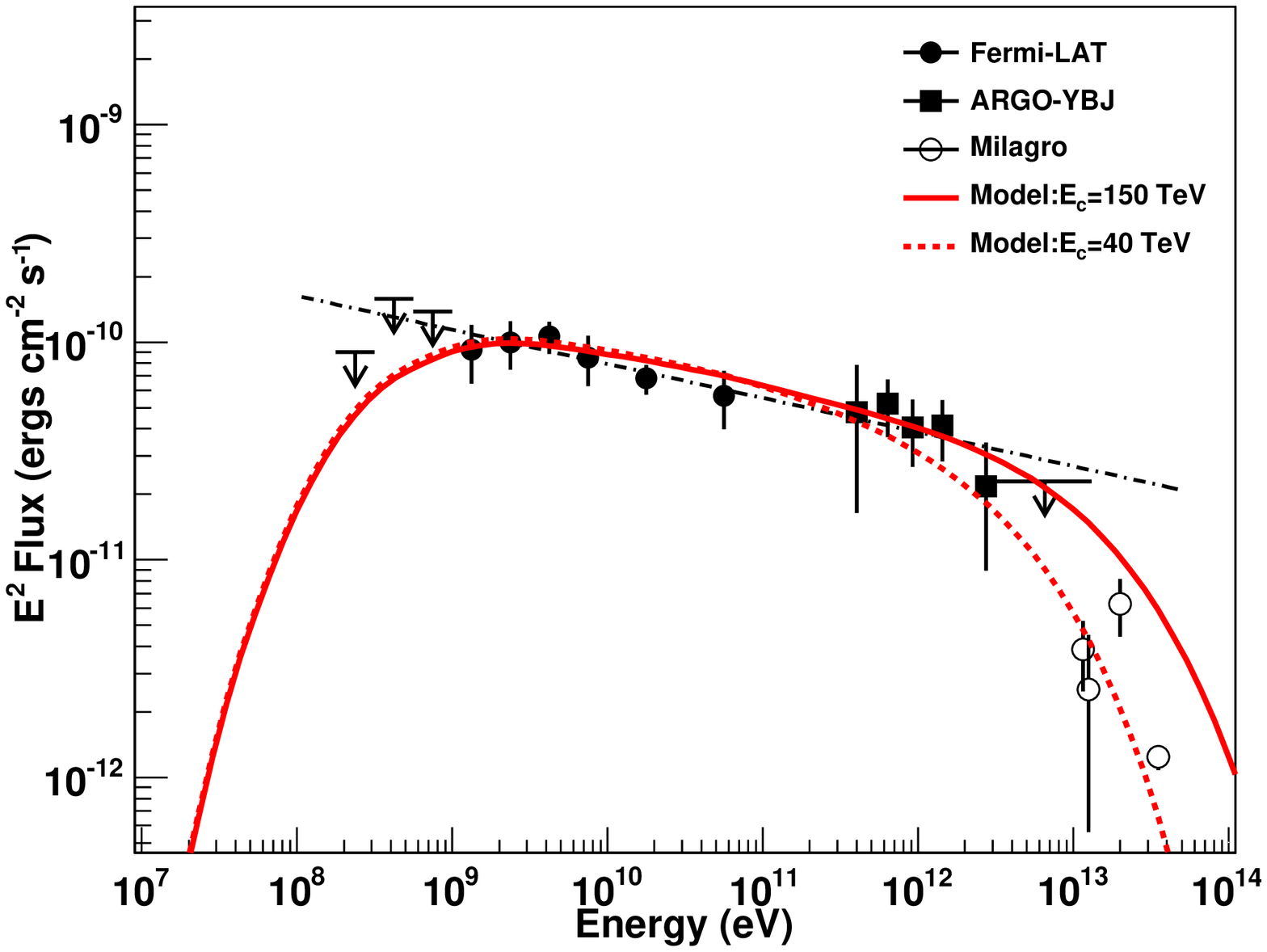} }
    \caption{Spectral energy distribution of the Cygnus Cocoon. Different markers stand for the spectra measured by different detectors. The arrows below 1 GeV show the upper limits obtained by Fermi-LAT \cite{ackermann11}. The points at 12, 20 and 35 TeV report the measurement of MGRO J2031+41 by Milagro \cite{abdo07,abdo09}. 
The lower data point at 12 TeV is the Milagro flux after the subtraction of the TeV J2032+4130 contribution \cite{ackermann11}. The thin blue solid line shows the best fit to the Fermi-LAT and ARGO-YBJ data using a simple power-law function. The thick red solid line is obtained by a hadronic model with a proton cutoff energy at 150 TeV, the dashed one refers to a proton cutoff energy of 40 TeV.}
  \label{fig:fig05}
        \end{minipage}\hfill
\end{figure}

The significance map around ARGO J2031+4157 as observed by ARGO-YBJ using events with N$_{pad}\geq$ 20 is shown in Fig. \ref{fig:fig04}. For comparison, the known TeV sources and the Cygnus Cocoon are marked in the figure. The sizes of markers indicate the 68\% containment size of the extension. The highest significance value in the figure is 6.1 s.d. corresponding to ARGO J2031+4157. The excess signals, larger than the angular resolution of the detector, almost fully fill the extension region of Cygnus Cocoon and MGRO J2031+41, indicating a similar extension size. At the top right corner of the extension there is a 4.3 s.d. peak corresponding to ARGO J2021+4038, which can be associated to VER J2019+407. 

The extension of ARGO J2031+4157 is estimated to be (1.5$^{+0.6}_{-0.9}$)$^{\circ}$. This result is consistent with the Cygnus Cocoon ($\sigma_{ext} = (2.0 \pm 0.2)^{\circ}$ \cite{ackermann11}) and the MGRO J2031+41 ($\sigma_{ext} = (1.8)^{\circ}$ \cite{abdo12}) sizes within the uncertainties.
An extension size of about 2$^{\circ}$ is larger than any identified Galactic TeV source. 
%
%
The photon spectrum can be described by a differential power-law, $dN/dE = (2.75\pm0.42)\times 10^{-11}\cdot (E/TeV)^{(-2.58\pm 0.23)}$ photons cm$^{-2}$ s$^{-1}$ TeV$^{-1}$, in energy between 200 GeV and 10 TeV. The flux is higher than in \cite{bartoli12b} since a large source region is used.

To derive the possible emission due to the Cygnus Cocoon, the contribution from the overlapping sources, i.e, TeV J2032+4130 and VER J2019+407, and nearby sources, i.e., VER J2016+372 and MGRO J2019+37, must be excluded.  After removing these contributions, the photon spectrum can be described by the following differential power law:
$dN/dE = (2.47\pm 0.42)\times 10^{-11}\cdot (E/TeV)^{(-2.62\pm 0.27)}$ photons cm$^{-2}$ s$^{-1}$ TeV$^{-1}$.
The integral flux above 1 TeV is (1.52$\pm$ 0.37)$\times$10$^{-11}$ photons cm$^{-2}$ s$^{-1}$, corresponding to 0.82$\pm$ 0.20 Crab units. 

The energy spectrum of the Fermi Cocoon measured by Fermi-LAT, ARGO-YBJ and Milagro is shown in Fig. \ref{fig:fig05}. The flux determined by ARGO-YBJ appears consistent with the extrapolation of the Fermi-LAT spectrum suggesting that the emission of ARGO J2031+4157 can be identified as the counterpart of Cygnus Cocoon at TeV energies. 
The combined spectrum of Fermi-LAT and ARGO-YBJ can be described by the following differential power law (thin solid line) $dN/dE = (3.46\pm 0.33)\times 10^{-9}\cdot (E/0.1TeV)^{(-2.16\pm 0.04)}$ photons cm$^{-2}$ s$^{-1}$ TeV$^{-1}$, suggesting the same origin for both GeV and TeV extended gamma-ray emission. Only statistical errors are shown, the systematic errors on the flux are estimated to be less than 30\% \cite{bartoli11b}.
The upper limits of Fermi-LAT and ARGO-YBJ indicate the presence of a slope change or cutoff below $\sim$1 GeV and above $\sim$10 TeV, respectively.

The angular size of about 2$^{\circ}$ places the Cygnus Cocoon among the most extended VHE gamma-ray sources. At a distance of 1.4 kpc, the observed angular extension corresponds to more than 50 pc, making the Cygnus Cocoon the largest identified Galactic TeV source. Such a large region can be related to different scenarios.
As discussed in \cite{ackermann11}, the favored scenario to explain the emission in the Fermi Cocoon is the injection of cosmic rays via acceleration from the collective action of multiple shocks from supernovae and the winds of massive stars, which form the Cygnus superbubble. Such superbubbles have been long advocated as cosmic ray factories, therefore the Cygnus Cocoon could be the first evidence supporting such hypothesis.

In order to test a possible hadronic origin of the gamma-ray emission through the $\pi^{0}$ decay, we considered inelastic collisions between accelerated protons and target gas. We assumed that the primary protons follow a power law with an index similar to the gamma spectrum and with an exponential cutoff at 150 TeV, as suggested in \cite{ackermann11} to describe CR acceleration by random stellar winds in the Cygnus superbubble. This energy is the maximum  proton cutoff allowed by the ARGO-YBJ upper limit. The resulting spectrum is shown in Fig. \ref{fig:fig05} by the thick solid red line. 
It is worth noting that the Milagro data are not described by this model and can be reconciled only with a cutoff of about 40 TeV (dashed red line) non consistent with the ARGO-YBJ results \cite{argo-cygcocoon}.

\section{Light component (p+He) spectrum of Cosmic Rays}

The CR light component (p+He) has been selected by ARGO-YBJ and its energy spectrum measured in the energy range (5 - 200) TeV. 
With this analysis for the first time a ground-based measurement of the CR spectrum overlaps data obtained with direct methods for more than one energy decade, thus providing a solid anchorage to the CR spectrum measurements carried out by EAS arrays in the knee region.

%
\begin{figure}
\centerline{\includegraphics[width=0.80\textwidth,clip]{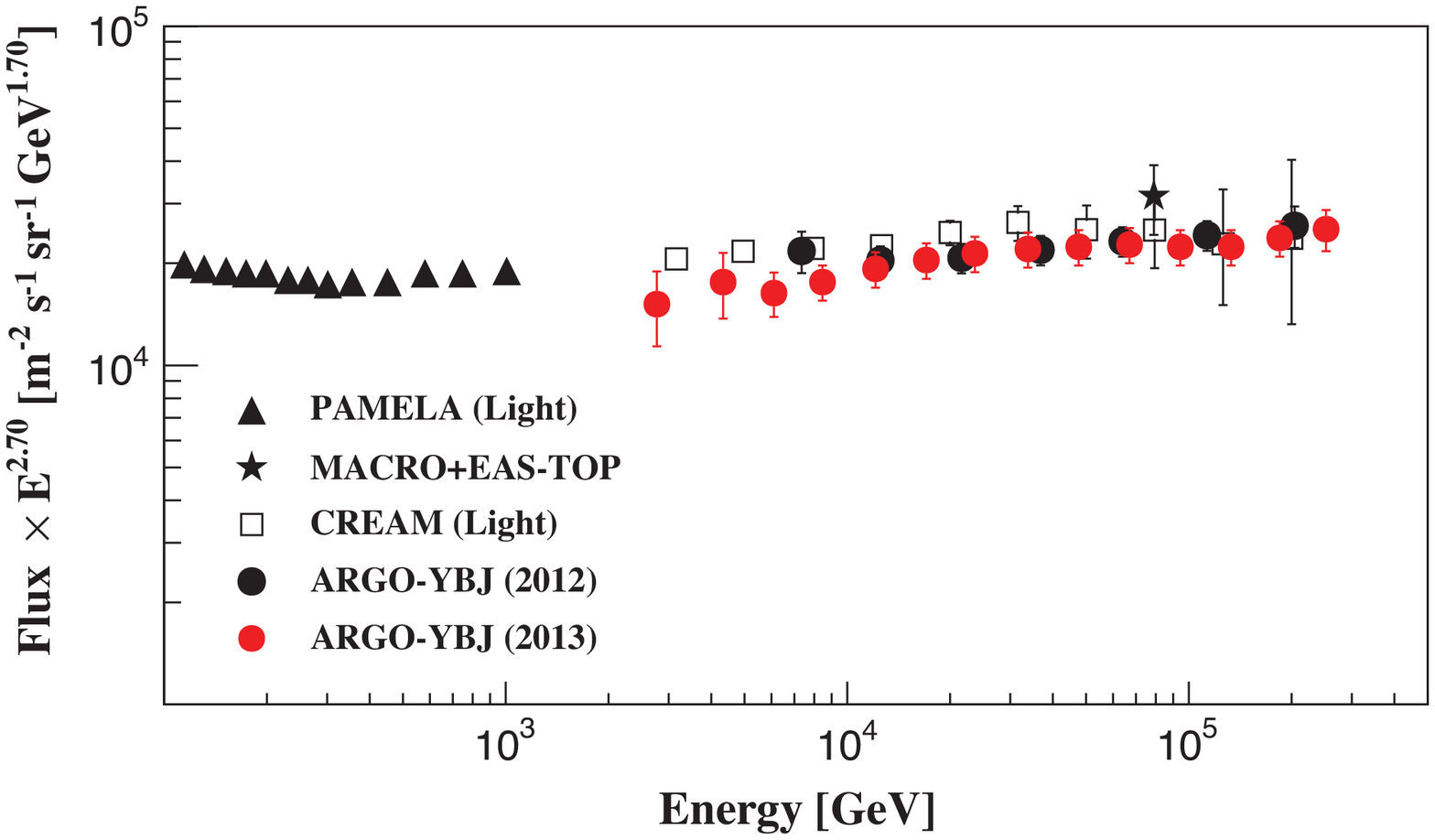} }
\caption{Light component (p+He) energy spectrum of primary CRs measured by ARGO-YBJ compared with other experimental results. The ARGO-YBJ 2012 data refer to the results published in \cite{bartoli12c} and the 2013 ones have been obtained with the full statistics.}
\label{fig:light_spectrum}       
\end{figure}
%
Requiring quasi-vertical showers ($\theta$ $<$ 30$^{\circ}$) and applying a selection criterion based on the particle density, a sample of events mainly induced by protons and helium nuclei, with shower core inside a fiducial area (with radius $\sim$28 m), has been selected. The contamination by heavier nuclei is found negligible. An unfolding technique based on the Bayesian approach has been applied to the strip multiplicity distribution in order to obtain the differential energy spectrum of the light component \cite{bartoli12c}. 
The spectrum measured by ARGO-YBJ is compared with other experimental results in Fig. \ref{fig:light_spectrum}.  
Systematic effects due to different hadronic models (Corsika 6.710 with QGSJet-II and SYBILL \cite{corsika}) and to the selection criteria do not exceed 10\%.
The ARGO-YBJ data agree remarkably well with the values obtained by adding up the p and He fluxes measured by CREAM both concerning the total intensities and the spectral index \cite{cream11}. The value of the spectral index of the power-law fit to the ARGO-YBJ data is -2.61$\pm$0.04, which should be compared with $\gamma_p$ = -2.66$\pm$0.02 and $\gamma_{He}$ = -2.58$\pm$0.02 obtained by CREAM.
The present analysis does not allow the determination of the individual p and He contribution to the measured flux, but the ARGO-YBJ data clearly exclude the RUNJOB results \cite{runjob}. 

This measurement has been extended to higher energies exploiting an hybrid measurement with a prototype of the future Wide Field of view Cherenkov Telescope Array (WFCTA) of the LHAASO project \cite{lhaaso}.
The telescope, located at the south-east corner of the ARGO-YBJ detector, about 78.9 m away from the center of the RPC array, is equipped with 16$\times$16 photomultipliers (PMTs), has a FOV of 14$^{\circ}\times$16$^{\circ}$ with a pixel size of approximately 1$^{\circ}$ \cite{nim2011}.

From December 2010 to February 2012, in a total exposure time of 728,000 seconds, the ARGO-YBJ/WFCTA system collected and reconstructed 8218 events above 100 TeV according to the following selection criteria: (1) reconstructed shower core position located well inside the ARGO-YBJ central carpet, excluding an outer region 2 m large; (2) more than 1000 fired pads on the central carpet; (3) more than 6 fired pixels in the PMT matrix; (4) a space angle between the incident direction of the shower and the telescope main axis less than 6$^{\circ}$.
This selection guarantees that the Cherenkov images are fully contained in the FOV, an angular resolution better than 0.2$^{\circ}$ and a shower core position resolution less than 2 m.

\begin{figure}
\begin{minipage}[ht]{.47\linewidth}
  \centerline{\includegraphics[width=\textwidth]{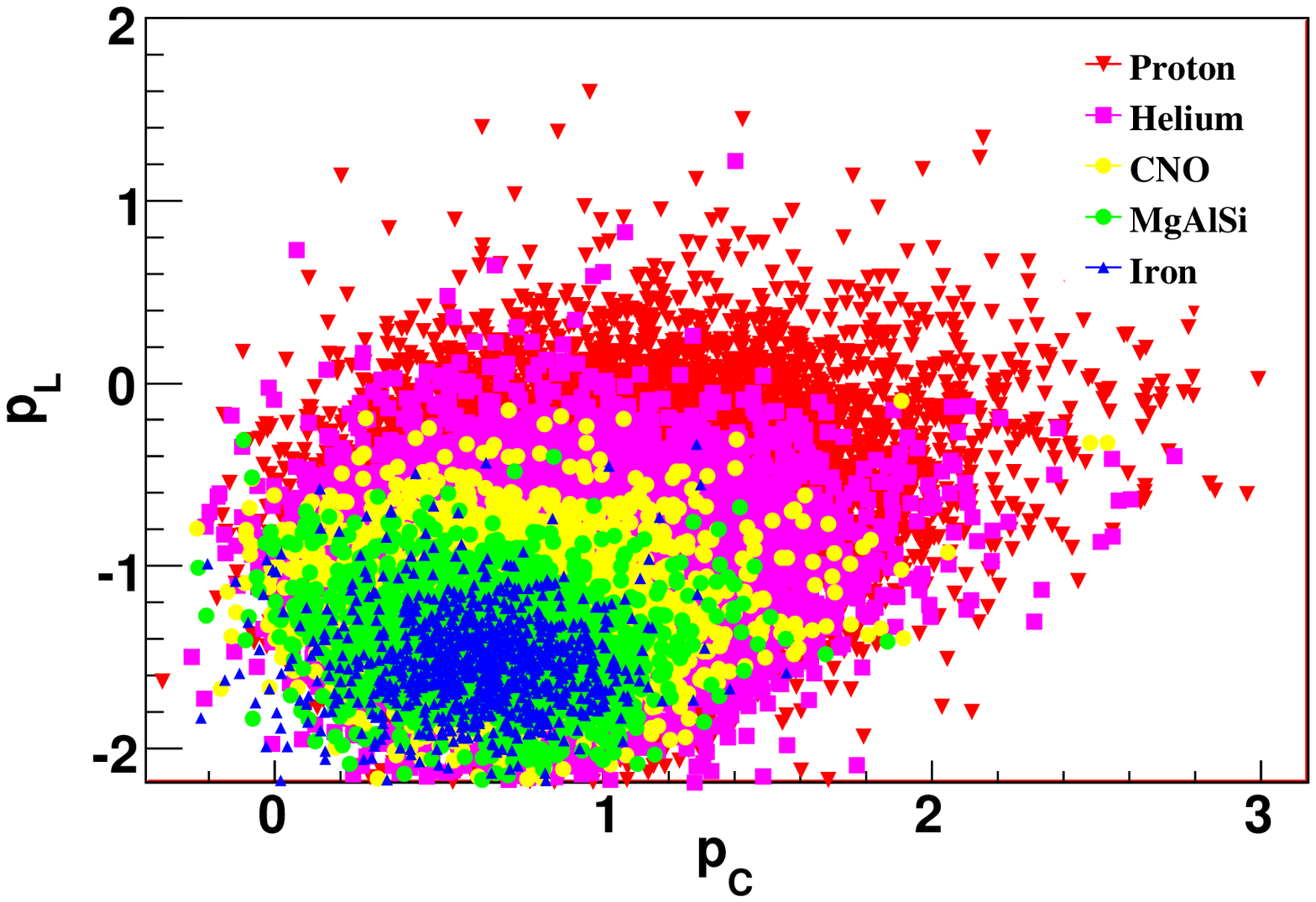}}
    \caption{Scatter plot of the parameters $p_C$ and $p_L$ for showers induced by different nuclei. The primary masses have been simulated in the same relative percentage.}
\label{fig:pl-pc}
\end{minipage}\hfill
\begin{minipage}[ht]{.47\linewidth}
  \centerline{\includegraphics[width=\textwidth]{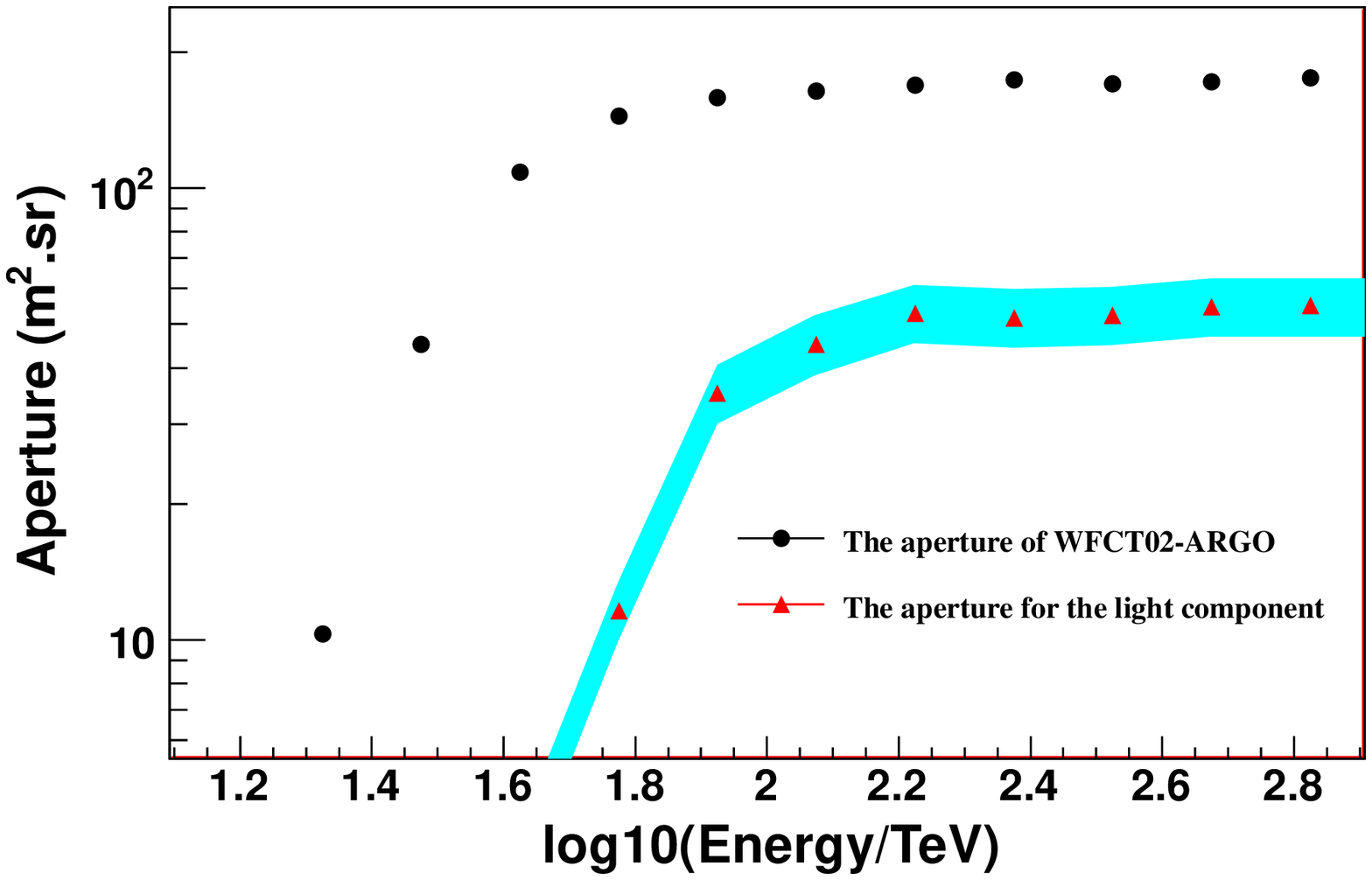} }
    \caption{The aperture of the hybrid experiment ARGO-YBJ/WFCTA. Filled circles refer to the all-particle, triangles to the selected light nuclei. The shaded area shows the systematic uncertainty.}
  \label{fig:wfcta-aperture}
        \end{minipage}\hfill
\end{figure}

According to the MC simulations, the largest number of particles N$_{max}$ recorded by a RPC in an given shower is a useful parameter to measure the particle density in the shower core region, i.e. within 3 m from the core position.
For a given energy, in showers induced by heavy nuclei N$_{max}$ is smaller than in showers induced by light particles. Therefore, N$_{max}$ is a parameter useful to select different primary masses.
In addition, N$_{max}$ is proportional to E$_{rec}^{1.44}$, where E$_{rec}$ is the shower primary energy reconstructed using the Cherenkov telescope.
We can define a new parameter p$_L$ = $log_{10} (N_{max}) - 1.44\cdot log_{10} (E_{rec}/TeV)$ by removing the energy dependence \cite{argo-wfcta}.

%
\begin{figure}
\centerline{\includegraphics[width=0.7\textwidth,clip]{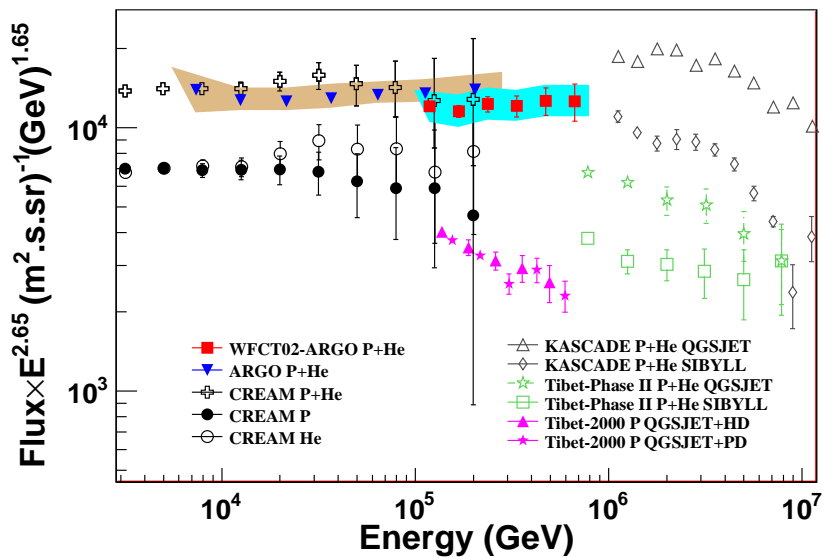} }
\caption{Light component (p+He) energy spectrum of primary CRs measured by ARGO-YBJ/WFCTA hybrid experiment (filled red squares) in the energy range 100 -- 700 TeV, compared with other experimental results. The ARGO-YBJ data at lower energies are published in \cite{bartoli12c}.}
\label{fig:wfcta-enspt}       
\end{figure}
%

The Cherenkov footprint of a shower can be described by the well-known Hillas parameters \cite{hillas85}, i.e. by the width and the length of the image. 
Older showers which develop higher in the atmosphere, such as iron-induced events, have Cherenkov images more stretched, i.e. narrower and longer, with respect to younger events due to light particles which develop deeper.
Therefore, the ratio between the length and the width (L/W) of the Cherenkov image is expected to be another good estimator of the primary elemental composition.

Elongated images can be produced, not only by different nuclei, but also by showers with the core position far away from the telescope, or by energetic showers, due to the elongation of the cascade processes in the atmosphere. 
Simulations show that the ratio of L/W is nearly proportional to the shower impact parameters R$_p$, the distance between the telescope and the core position, which must be accurately measured.
An accurate determination of the shower geometry is crucial for the energy measurement. In fact, the number of photoelectrons collected in the image recorded by the Cherenkov telescope N$_{pe}$ varies dramatically with the impact parameter R$_p$, because of the rapid falling off of the lateral distribution of the Cherenkov light. 
Only an accurate measurement of the shower impact parameters R$_p$, and a good reconstruction of the primary energy allow to disentangle different effects.
A shower core position resolution better than 2 m and an angular resolution better than 0.2$^{\circ}$, due to the high-granularity of the ARGO-YBJ full coverage carpet, allow to reconstruct the shower primary energy with a resolution of 25\%, by using the total number of photoelectrons N$_{pe}$. The uncertainty in absolute energy scale is estimated about 10\%.

Therefore, in order to select the different masses we can define another new parameter p$_C$ = $L/W - 0.0091\cdot(R_p/1\>m) - 0.14\cdot log_{10}(E_{rec} /TeV)$ by removing both the effects due to the shower distance and to the energy.

The values of these parameters for showers induced by different nuclei are shown in the Fig. \ref{fig:pl-pc}. The events have been generated assuming a -2.7 spectral index in the energy range 10 TeV -- 10 PeV for all the five mass groups (p, He, CNO, MgSi, Fe) investigated. The primary masses have been simulated in the same relative percentage.
As can be seen from the figure, a suitable selection in the p$_L$ -- p$_C$ space allows to pick out a light composition sample with high purity. In fact, by cutting off the concentrated heavy cluster in the lower-left region in the scatter plot, i.e. p$_L\leqslant$ -0.91 and p$_C \leqslant$ 1.3, the contamination of nuclei heavier than He is less than 5\%. 
About 30\% of H and He survives the selection criteria.

The aperture of the ARGO-YBJ/WFCTA system has been estimated using the Horandel model for the primary spectrum \cite{horandel} and the Corsika QGSJETII-03/GHEISHA codes \cite{corsika} to describe the hadronic interactions.
Its value, $\sim$170 m$^2$sr above 100 TeV, shrinks to $\sim$50 m$^2$sr after the selection of the (p+He) component (see Fig. \ref{fig:wfcta-aperture}).

In the sample of 8218 events recorded above 100 TeV by the hybrid system, 1392 showers are selected in the (p+He) sub-sample.
The light component energy spectrum measured by the ARGO-YBJ/WFCTA hybrid system is shown in the Fig. \ref{fig:wfcta-enspt} by the filled red squares. A systematic uncertainty in the absolute flux of 15\% is shown by the shaded area. The error bars show the statistical errors only.
The spectrum can be described by a power law with a spectral index of -2.63 $\pm$ 0.06. The absolute flux at 400 TeV is (1.79$\pm$0.16)$\times$10$^{-11}$ GeV$^{-1}$ m$^{-2}$ sr$^{-1}$ s$^{-1}$. 
This result is consistent for what concern spectral index and absolute flux with the measurements carried out by ARGO-YBJ below 200 TeV and by CREAM. The flux difference is about 10\% and can be explained with a difference in the experiments energy scale of $\pm$3.5\%. 

In conclusion, the CR light component primary energy spectrum can be described by a single power-law up to about 600 TeV.

\section{Cosmic Ray Anisotropy}

The CR arrival direction distribution and its anisotropy has been a long-standing problem ever since the 1930s. In fact, the study of the anisotropy is a powerful tool to investigate the acceleration and propagation mechanism determining the CR world as we know it. 

The anisotropy in the CR arrival direction distribution have been observed by different experiments with increasing sensitivity and details at different angular scales.
Current experimental results show that the main features of the anisotropy are uniform in the energy range (10$^{11}$ - 10$^{14}$ eV), both with respect to amplitude (10$^{-4}$ - 10$^{-3}$) and phase ((0 - 4) hr).
The existence of two distinct broad regions, one showing an excess of CRs (called ``tail-in''), distributed around 40$^{\circ}$ to 90$^{\circ}$ in R.A., the other a deficit (the ``loss cone''), distributed around 150$^{\circ}$ to 240$^{\circ}$ in R.A., has been clearly observed (for a review see, for example, \cite{disciascio13}).

The origin of the CR anisotropy is still unknown.
Unlike predictions from diffusion models, the CR arrival distribution in sidereal time was never found to be purely dipolar. 
Even 2 harmonics were necessary to properly describe the R.A. profiles, showing that the CR intensity has quite a complicated structure unaccountable simply by kinetic models.

In the last years Tibet AS$\gamma$ \cite{tibetasg}, Milagro \cite{abdo08} and ARGO-YBJ \cite{bartoli13b} reported evidence of the existence of a medium angular scale anisotropy contained in the tail-in region. The observation of similar small scale anisotropies has been claimed also by the ICECUBE experiment \cite{abbasi11} in the southern hemisphere.

\begin{figure}
\centerline{\includegraphics[width=0.7\textwidth,clip]{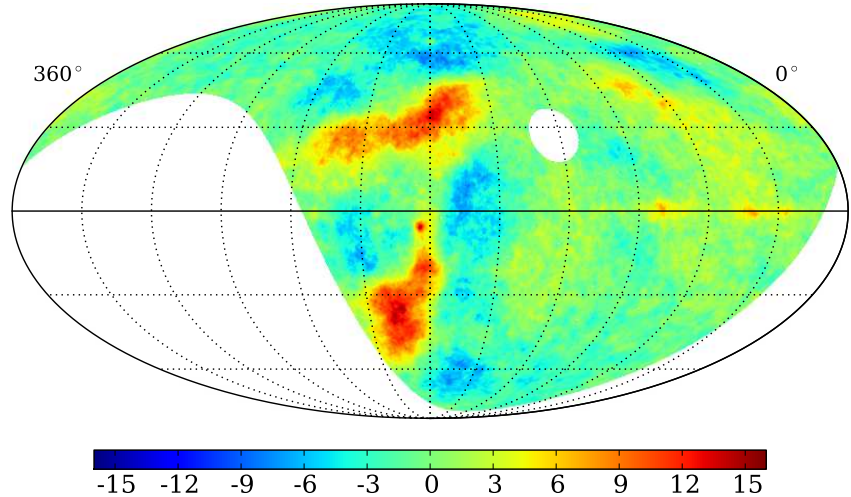}}
\caption[h]{ARGO-YBJ sky-map in galactic coordinates. The color scale gives the statistical significance of the observation in s.d. .  The map has been smoothed with an angle given by the point spread function of the detector for CR-induced showers.The map center points towards the galactic Anti-Center.}
\label{fig:fig10}
\end{figure}

In Fig. \ref{fig:fig10} the ARGO-YBJ sky map in galactic coordinates as obtained with 4.5 years data is shown. The map center points towards the galactic Anti-Center. 
Data have been recorded in 1587 days out of 1656, for a total observation time of 33012 hrs ($86.7\%$ duty-cycle). A selection of high-quality data reduced the data-set to 1571 days. 
The zenith angle cut ($\theta\leq$ 50$^{\circ}$) selects the declination region $\delta\sim$ -20$^{\circ}\div$ 80$^{\circ}$.
According to the simulation, the median energy of the isotropic CR proton flux is E$_p^{50}\approx$1.8 TeV (mode energy $\approx$0.7 TeV).
No gamma/hadron discrimination algorithms have been applied to the data. Therefore, the sky map is filled with all CRs possibly including photons, without any discrimination.

In spite of the fact that the bulk of SNR, pulsars and other potential CR sources are in the Inner Galaxy surrounding the Galactic Centre, the excess of CR is observed in the opposite, Anti-Centre direction. As stressed in \cite{erlykin13}, the fact that the observed excesses are in the Northern and in the Southern Galactic Hemisphere, favors the conclusion that the CR at TeV energies originate in sources whose directions span a large range of Galactic latitudes.

The right side of the map is full of few-degree excesses not compatible with random fluctuations (the statistical significance is up to 7 s.d.). The observation of these structures is reported by ARGO-YBJ for the first time.

So far, no theory of CRs in the Galaxy exists which is able to explain both large scale and few degrees anisotropies leaving the standard model of CRs and that of the local galactic magnetic field unchanged at the same time. 

\section{Conclusions}

The ARGO-YBJ detector exploiting the full coverage approach and the high segmentation of the readout is imaging the front of atmospheric showers with unprecedented resolution and detail in the wide TeV - PeV energy range. 

Since November 2007 to January 2013 the ARGO-YBJ experiment monitored with high duty cycle ($\sim$86$\%$) the northern sky at TeV photon energies. With a cumulative sensitivity ranging from 0.24 to $\sim$1 Crab units, depending on the declination, six sources have been observed with a statistical significance greater than 5 s.d. in the declination band from -10$^{\circ}$ to 70$^{\circ}$. 

A detailed study of the Cygnus region support the identification of the ARGO J2031+4157 source as the TeV counterpart of the so-called Cygnus Cocoon, observed by Fermi-LAT in the energy region 1--100 GeV, suggesting the same origin for both GeV and TeV gamma-ray emissions. 

In addition, the CR light component (p+He) energy spectrum has been measured up to about 600 TeV exploting for the first time a hybrid system with ARGO-YBJ and a wide field of view Cherenkov telescope.
We found that the light component spectrum can be described by a single power-law up to about 600 TeV.

Finally, the observation for the first time of many few-degree CR excesses in the Northern sky favors the conclusion that the CR at TeV energies originate in sources whose directions span a large range of Galactic latitudes.





\bibliographystyle{elsarticle-num}
\bibliography{<your-bib-database>}



\end{document}